\documentclass[amsmath,amssymb,aps,prfluids,unsortedaddress,superscriptaddress]{revtex4-2}
\usepackage{savesym} 

\usepackage{times} 
\usepackage{setspace} 
\usepackage{geometry} 
\usepackage[colorlinks=true,
            linkcolor=black,
            urlcolor=red,
            citecolor=blue]{hyperref}
\usepackage[dvipsnames]{xcolor}

\usepackage{graphicx} 
\usepackage[normal,labelsep=colon,skip=8pt,labelfont=bf]{caption}
\usepackage{wrapfig} 

\usepackage{multirow}
\usepackage{makecell}

\usepackage{amsmath} 
\usepackage{amssymb} 
\usepackage{amsthm} 
\usepackage{wasysym} 

\usepackage{comment}

\begin{document}
\title{Consistency requirement of data-driven subgrid-scale modeling in large-eddy simulation}

\author{Xinyi Huang}
\email{xinyih@brown.edu}
\affiliation{Graduate Aerospace Laboratories, California Institute of Technology, Pasadena, CA, 91125, USA}
\affiliation{Department of Earth, Environmental \& Planetary Sciences, Brown University, Providence, RI, 02906, USA}
\author{Sze Chai Leung}
\affiliation{Department of Mechanical and Civil Engineering, California Institute of Technology, Pasadena, CA, 91125, USA}
\author{H. Jane Bae}
\affiliation{Graduate Aerospace Laboratories, California Institute of Technology, Pasadena, CA, 91125, USA}

\begin{abstract}

Data-driven subgrid-scale (SGS) modeling in the large-eddy simulations (LES) suffers from the inconsistency between the \textit{a priori} tests and the \textit{a posteriori} tests, which make training accurate SGS models a difficult task. 
We study the difference in filtered high-fidelity data and LES to identify the numerical deviation between the two cases, which is a combined impact of commutation error, numerical errors, and error coupling. 
The impact of the numerical deviation is examined through two SGS model formulations: the eddy-viscosity and the complex nonlinear models. 
By incorporating numerical deviations into model training, we enhance consistency, stabilize simulations, and improve predictions of energy spectra in \textit{a posteriori} tests. Our findings highlight that data-driven methods introduce significant nonlinearity and equation coupling, exacerbating inconsistencies compared to non-data-driven approaches. 
Finally, while the impact of the numerical deviation can be generalized, achieving accurate model predictions necessitates a physically grounded model form and an optimal filter width. 

\end{abstract}

\maketitle

\section{Introduction}
\label{sec:Intro}



Large-eddy simulation (LES) is a computational approach that resolves the large, energy-containing structures of turbulence while modeling the effects of smaller-scale motions, significantly reducing the computational cost compared to direct numerical simulation (DNS). Over the past few decades, LES has become a powerful tool for investigating a wide range of flows—from canonical configurations to complex real-world scenarios—offering rich physical insights \cite{goc2021large,okaze2021large,zahn2024setting,huang2025characteristics,xiao2021large}.
To extend the applicability of subgrid-scale (SGS) models to practical flows, it is essential to incorporate physical understanding of complex phenomena such as flow separation, pressure gradients, surface roughness, stratification, and compressibility \cite{pino2000subgrid,stoll2006dynamic,agrawal2022non,unnikrishnan2021subgrid}. However, developing SGS models for such flows is challenging due to factors including anisotropy, significant numerical discretization effects, and spatial inhomogeneity in resolution \cite{moser2021statistical}.
Extracting relevant physical features for model development in these regimes often demands considerable effort. In this context, recent advances in data-driven modeling offer promising opportunities to streamline and enhance SGS model development.

In recent years, data-driven methods have gained traction in SGS modeling due to their potential to uncover complex relationships between SGS terms and resolved quantities using high-fidelity data \cite{gamahara2017searching,park2021toward}. However, most developments have focused on canonical flows \cite{zhou2019subgrid,kang2023neural,maejima2024coarse}, with limited success in more complex flow scenarios \cite{macart2021embedded,meng2023artificial,kim2024large}.
Several factors contribute to the slow progress in applying data-driven SGS models to complex flows. First, these models face additional challenges in practical applications, such as numerical instability during deployment \cite{park2021toward}, limited performance gains over traditional models \cite{zhou2019subgrid,meng2023artificial}, and dependence on case-specific global information \cite{macart2021embedded}. Even in canonical settings, data-driven models that appear highly accurate in training (\textit{a priori}) tests can become unstable in simulations, often requiring artificial backscatter clipping and still underperforming classical models in \textit{a posteriori} evaluations.
Second, the ``black-box'' nature of many data-driven approaches makes it difficult to incorporate known physical principles and constraints—an advantage of traditional models. For instance, enforcing Galilean invariance is generally believed to enhance model generalizability \cite{prakash2022invariant}. Yet, due to interactions between numerical discretization and the online training process, enforcing seemingly appropriate constraints such as symmetry of the SGS stress tensor can sometimes degrade model performance significantly \cite{macart2021embedded}.

A major limitation of data-driven SGS models lies in the inconsistency between their \textit{a priori} and \textit{a posteriori} performance. During model development, i.e., in \textit{a priori} testing, training and validation data are typically generated by filtering high-fidelity simulations. However, in \textit{a posteriori} tests, the model is deployed in LES, which evolve under different governing equations than those used to produce filtered DNS data. This mismatch arises from commutation errors between discretization and differentiation operators, as well as the coarse spatial resolution inherent to LES \cite{bae2022numerical}. As a result, numerical derivatives of filtered quantities in LES often differ significantly from the filtered derivatives computed from high-resolution data. For example, given a quantity $\phi$, unless specific filter and numerical derivative operators are chosen,
\begin{equation}
\label{eq:comm-err}
\overline{\frac{\Delta \phi}{\Delta x_i}} \neq \frac{\Delta \overline{\phi}}{\Delta x_i}, \quad \overline{\frac{\Delta \phi}{\Delta t}} \neq \frac{\Delta \overline{\phi}}{\Delta t},  
\end{equation} 
where the filtered quantities are represented by $\overline{(\cdot)}$. 
Note that after numerical discretization, the aforementioned two numerical derivatives are not necessarily performed on the same grid resolution, leading to additional numerical discrepancies even when using the same numerical scheme for both the DNS and LES. 
By considering all numerical discrepancies, the LES equations for the incompressible Navier-Stokes equations consistent with filtered DNS are 
\begin{equation}
\frac{\partial \overline{u}_i}{\partial t} + \frac{\partial {\overline{u}_i \overline{u}_j}}{\partial x_j} = - \frac{1}{\rho}\frac{\partial \overline{p}}{\partial x_i} + \nu \frac{\partial^2 \overline{u}_i}{\partial x_j \partial x_j} - \frac{\partial \tau_{ij}}{\partial x_j} + \delta_i^{\textrm{mom}},\quad\frac{\partial \overline{u}_i}{\partial x_i} + \delta^{\textrm{div}}=0, 
\label{eq:NS-delta}
\end{equation} 
where $x_i$ is the Cartesian coordinate with $i\in \{1,2,3\}$, and $u_i$ is the corresponding velocity component in the respective direction. 
The quantity given by $\nu$ is the kinematic viscosity, $p$ is the pressure, and $\rho$ is the fluid density. 
The SGS stress tensor is defined as $\tau_{ij} = \overline{u_i u_j} - {\overline{u}_i \overline{u}_j}$. 
The extra terms $\delta_i^{\textrm{mom}}$ and $\delta^{\textrm{div}}$ are the numerical deviation that differentiate the \textit{a priori} tests from the \textit{a posteriori} tests. 
Consequently, the actual values are dependent on the specific choice of the filter and the numerical scheme. 


The inconsistency between \textit{a priori} and \textit{a posteriori} results is not unique to data-driven SGS model development; it has also been observed in conventional models \cite{piomelli1988model}. In practice, this inconsistency can be mitigated through appropriate choices in the filtering operator and the use of explicit filtering—both aimed at aligning the modeling assumptions used during training with those encountered during LES deployment. For instance, selecting an appropriate filter type can significantly improve \textit{a posteriori} performance. As demonstrated in Ref. \cite{piomelli1988model}, combinations such as the mixed model with a Gaussian filter and the Smagorinsky model with a spectral cutoff filter are considered consistent and effective.
Moreover, increasing the ratio between the filter width and the grid spacing (i.e., using a larger filter-grid ratio) helps reduce numerical error in LES, as it increases the relative contribution of the subfilter-scale stresses \cite{ghosal1996analysis,chow2003further}. For nonlinear numerical schemes or non-uniform grid spacing, the construction of discrete filters must follow specific design rules to minimize commutation errors \cite{vasilyev1998general,marsden2002construction,bose2010grid}. In this context, explicit filtering—where an additional filter is applied to the nonlinear terms in the Navier–Stokes equations—has been shown to enhance \textit{a posteriori} model performance \cite{lund2003use,carati2001modelling,winckelmans2001explicit}.

Several strategies have been proposed to mitigate the issue of inconsistency, including the use of commutative filters, increasing the filter size, and applying explicit filtering techniques. While these approaches have achieved partial success in improving the consistency between \textit{a priori} and \textit{a posteriori} modeling, they do not fully eliminate the numerical discrepancies. These discrepancies are highly dependent on the specific numerical scheme employed, making them difficult to characterize or correct \textit{a priori}.
The challenge is further compounded by the strong coupling of various error sources during the evolution of the governing equations, which leads to the accumulation of numerical and modeling errors over time \cite{bae2022numerical}. This issue is particularly severe in data-driven modeling, where nonlinear models are highly sensitive to deviations from the training data distribution. As the simulation evolves, accumulated errors can drive the solution far from the region of state space for which the model was trained, leading to instability and poor predictive performance \cite{sarghini2003neural}.

To mitigate the issue of accumulating errors, early efforts based on optimal estimation theory aimed to approximate an ``ideal'' LES using functions of the resolved LES fields \cite{langford1999optimal}. In this context, the ideal LES corresponds to filtered high-resolution (e.g., DNS) data, and the goal of optimal formulations is to reproduce the key statistics of such filtered fields. However, substantial relative errors have been observed across a broad range of spectral scales. Capturing high-wavenumber behavior accurately requires access to global information and high-order multipoint correlation data—resources that are typically unavailable during simulations—and the approach has been largely restricted to homogeneous, isotropic turbulence \cite{langford2004optimal,moser2009theoretically}.
An alternative approach is to directly model the numerical discrepancy in the LES governing equations \eqref{eq:NS-delta} during the data-driven training phase \cite{beck2019deep,kurz2023deep,beck2023toward}. While this strategy explicitly targets the numerical deviation, models trained in this way often suffer from numerical instability in \textit{a posteriori} tests, again due to the accumulation of error over time \cite{beck2019deep}. More recent online learning approaches, such as reinforcement learning, have demonstrated improved numerical stability; however, they require access to non-local flow information and remain highly case-specific, limiting their general applicability \cite{kurz2023deep,beck2023toward}.


Building on previous efforts to improve model consistency, we examine the restrictions that should be imposed on data-driven SGS model development. In practice, SGS models typically rely on locally resolved flow quantities to facilitate generalization across different geometries. While incorporating nonlocal information has shown promise—for example, in dynamic procedures for traditional models \cite{meneveau2000scale}, and through convolutional neural networks (CNNs) \cite{kurz2023deep} or recurrent neural networks (RNNs) \cite{kurz2022machine} in data-driven frameworks—such approaches often introduce additional challenges related to spatial grid structure or temporal schemes.
To maintain flexibility and avoid these complications, we focus exclusively on local input features, eliminating dependencies on grid structure or time history. Simultaneously, we enforce Galilean invariance, ensuring that the model is invariant under translations, rotations, and uniform motions. This constraint guides our model formulation, which is constructed from scalar and tensor invariants of the local velocity gradient, following the classical framework proposed by Pope \cite{pope1975more} and later used in complex nonlinear models \cite{lund1993parameterization}.
In the context of data-driven modeling, imposing physical constraints such as Galilean invariance enhances generalizability and interpretability \cite{prakash2022invariant}. Moreover, incorporating a richer set of tensor invariants helps compensate for the lack of a local equilibrium assumption and enables more expressive model forms \cite{inagaki2023analysis}. Based on these considerations, we explore two representative classes of SGS models in this study: an eddy-viscosity model and a complex nonlinear model.

In this work, we investigate the consistency between \textit{a priori} and \textit{a posteriori} tests in data-driven SGS modeling. Section \ref{sec:sim} describes the DNS-aided LES framework and the procedure for generating training data for the data-driven models. In Section \ref{sec:results}, we analyze how numerical deviations influence both \textit{a priori} training accuracy and \textit{a posteriori} simulation performance. The generalizability of these findings is further discussed in Section \ref{sec:discuss}. Finally, the main conclusions are summarized in Section \ref{sec:conclusion}.

\section{Simulation details and data-driven methods}
\label{sec:sim}


\subsection{Direct numerical simulation aided large-eddy simulation}
\label{sub:DNSaidLES}



\begin{figure}
    \centering
    \includegraphics[width=0.9\linewidth]{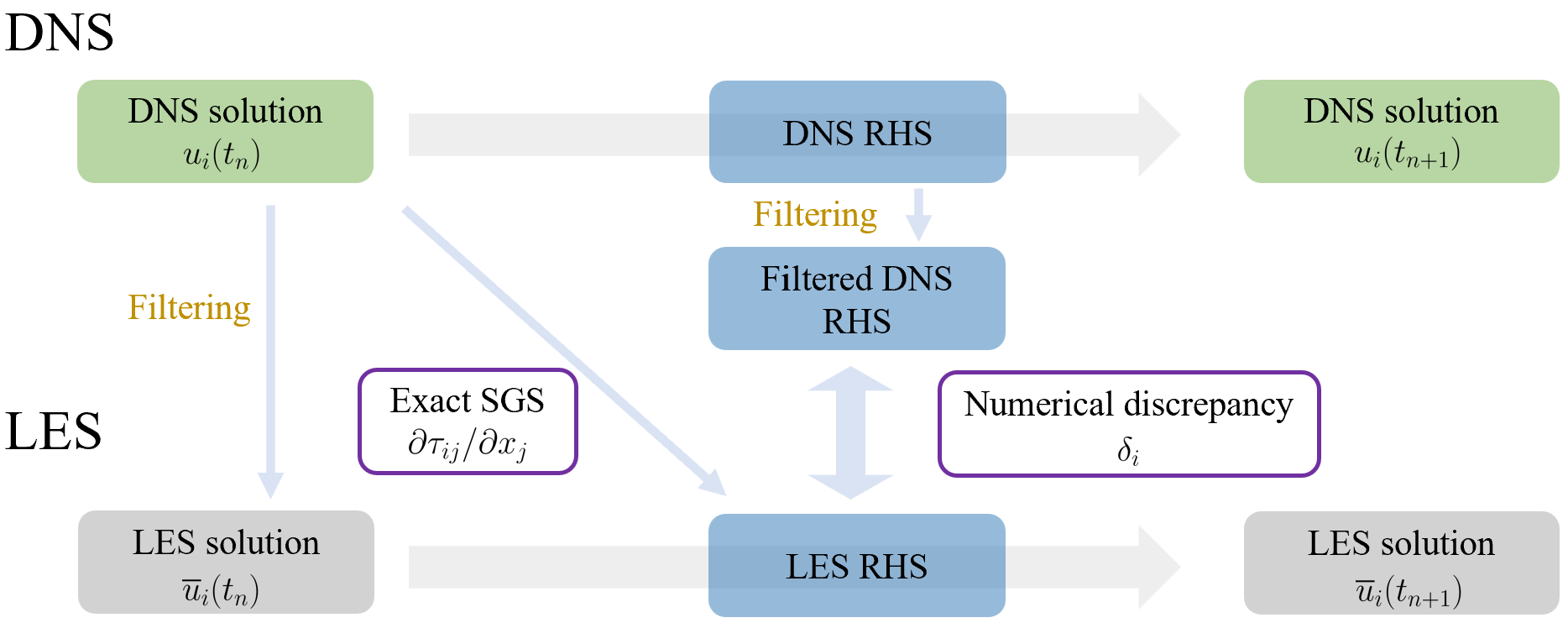}
    \caption{A sketch of the DNS-aided LES evolving from time step $t_n$ to time step $t_{n+1}$. }
    \label{fig:intro-NumErr}
\end{figure}

To accurately quantify the numerical deviation in Eq. \eqref{eq:NS-delta}, we employ a DNS-aided LES framework, in which filtered DNS and LES are run side-by-side to directly evaluate the discrepancies between them \cite{bae2022numerical}. This setup is illustrated schematically in Fig. \ref{fig:intro-NumErr}, which shows the evolution from time step $t_n$ to $t_{n+1}$.
At each time step, the DNS flow field is filtered to generate the exact subgrid-scale (SGS) stress divergence $\partial \tau_{ij}/\partial x_j$. By comparing the right-hand side (RHS) terms of the filtered DNS and the LES, highlighted as blue boxes in the figure, we can isolate the numerical deviation term $\delta_i$. With this deviation accounted for, the LES can be evolved such that its solution remains consistent with that of the filtered DNS.
All simulations are performed using an in-house code that employs a staggered-grid, second-order finite-difference scheme in space, and advances the solution in time using a third-order Runge–Kutta method. The DNS solves the discrete incompressible Navier–Stokes equations, with incompressibility enforced via the projection method.

In the present DNS-aided LES framework, both the DNS and LES are advanced in time using the same time integration scheme and time step, for simplicity. 
All filtered quantities in Eq. \eqref{eq:NS-delta} are computed directly from the DNS solution. 
Notably, after filtering, the resulting fields, including the SGS stress tensor $\tau_{ij}$, remain defined on the DNS grid. 
To be used in the LES, these filtered quantities are interpolated and downsampled to the LES grid, depending on the resolution and numerical scheme.
This interpolation process can be interpreted as an additional filtering operation imposed by the underlying grid structure. 
In this study, linear interpolation is employed on a staggered grid. 
{
Derivative operators are applied after the interpolation to the LES grid.
Therefore, the created training dataset aligns with the LES simulations of the \textit{a posteriori} tests, where the derivatives are always computed on the LES grid. 
Conversely, DNS quantities can also be filtered and interpolated after the derivatives are taken, but this is not adopted in this study.} 
In either case, when the filtering and derivative operators do not commute, the LES equations deviate from the filtered DNS equations. 
For this study, we neglect the commutation error in the time derivative term.
For simplicity, {
due to the existence of the projection step in both the DNS and LES, we regard any numerical inconsistency in the continuity equation, denoted $\delta^{\textrm{div}}$, as directly absorbed during the projection step.} 
We therefore focus solely on the momentum equation and drop the superscript ``mom'' from $\delta^{\textrm{mom}}$ without loss of clarity. 
{
The numerical deviation $\delta_i$ is a combined discrepancy in both the momentum equation and the continuity equation.}
The LES equation with the numerical deviation can be written as
\begin{equation}
\frac{\partial \overline{u}_i}{\partial t} + \frac{\partial {\overline{u}_i \overline{u}_j}}{\partial x_j} = - \frac{1}{\rho}\frac{\partial \overline{p}}{\partial x_i} + \nu \frac{\partial^2 \overline{u}_i}{\partial x_j \partial x_j} - \frac{\partial \tau_{ij}}{\partial x_j} + \delta_i,\quad\frac{\partial \overline{u}_i}{\partial x_i}=0.
\label{eq:NS-delta-momonly}
\end{equation} 

For further analysis, we can also derive the budget equation of the resolved kinetic energy $K_r=\overline{u}_i\overline{u}_i/2$ to be 
\begin{align}
    \frac{\partial K_r}{\partial t} = & \underbrace{- \frac{\partial K_r\overline{u}_j }{\partial x_j}}_{\mathcal{D}_\textrm{resolved}}
    \underbrace{-\frac{1}{\rho}\frac{\partial \overline{u}_i\overline{p}}{\partial x_i}}_{\mathcal{D}_{\textrm{pressure}}} 
    + \underbrace{\nu\frac{\partial^2 K_r}{\partial x_j \partial x_j}}_{\mathcal{D}_{\textrm{viscous}}} 
    - \underbrace{\frac{\partial \overline{u}_i\tau_{ij}}{\partial x_j} }_{\mathcal{D}_{\textrm{SGS}}} \nonumber\\
    & \qquad\qquad\qquad\quad
    - \underbrace{\nu\frac{\partial \overline{u}_i}{\partial x_j}\frac{\partial \overline{u}_i}{\partial x_j}}_{\varepsilon_{\textrm{viscous}}} 
    + \underbrace{\tau_{ij}\frac{\partial \overline{u}_i}{\partial x_j} }_{\varepsilon_{\textrm{SGS}}}
    + \underbrace{\overline{u}_i\delta_i^{\textrm{mom}}}_{\varepsilon_{\textrm{numerical}}}. 
    \label{eq:NS-ke}
\end{align}
The right-hand-side budget terms consist of the convection due to the resolved velocity $\mathcal{D}_\textrm{resolved}$, the pressure transport term $\mathcal{D}_\textrm{pressure}$, the SGS transport term $\mathcal{D}_\textrm{SGS}$, the viscous dissipation term $\varepsilon_{\textrm{viscous}}$, the SGS dissipation term $\varepsilon_{\textrm{SGS}}$, and the numerical dissipation term $\varepsilon_{\textrm{numerical}}$. 

The choice of filtering operator has a significant impact on the inconsistency between DNS and LES solutions, as demonstrated in \cite{huang2024consistent}. One way to reduce the relative importance of the numerical deviation is to employ a larger filter size. However, increasing the filter size also diminishes the resolved energy in the smaller turbulence scales and substantially raises the uncertainty in modeling the unresolved scales. This trade-off and its implications is discussed in detail in Sec. \ref{sub:generalize}.
In this study, we focus on the standard Gaussian filter, where the filtering operator $\overline{(\cdot)}$ is defined as 
\begin{equation}
    \overline{\phi}(\boldsymbol{x},t) = \int G(\mathbf{r})\phi(\boldsymbol{x-r},t) \textrm{d} \mathbf{r}, \quad G(\mathbf{r})=\frac{1}{\sqrt{2\pi}\sigma}e^{-\frac{|\mathbf{r}|^2}{2\sigma^2}}. 
\end{equation}
We utilize multiple filter widths, $\sigma/\Delta_{\textrm{LES}}$ of 0.5, 1.0, 2.0, 4.0, where $\sigma/\Delta_{\textrm{LES}}=2.0$ is the baseline filter width. 

DNS-aided LES is carried out for a forced homogeneous isotropic flow at Reynolds number based on Taylor microscale of $Re_{\lambda}=\sqrt{2k/3}\lambda/\nu=180$, where $k$ is the turbulent kinetic energy, $\varepsilon$ is the dissipation rate and $\lambda=\sqrt{10k/\varepsilon}$. 
The Kolmogorov length scale for this flow is $\eta=\sqrt[4]{\nu^3/\varepsilon}$, and the Kolmogorov time scale is given by $\tau_\eta=\sqrt{\nu/\varepsilon}$. 
A linear forcing is used to provide stationary turbulence as proposed by \cite{lundgren2003linearly}. 
We use a periodic box of size $(2\pi L_x)^3$ with a resolution of $256^3$ for the DNS, where $L_x$ is the characteristic length for the domain, which corresponds to $\Delta_{\textrm{DNS}} \approx 5.1\eta$. 
The time step is given by $\Delta t_{\textrm{DNS}} = 0.013\tau_\eta$. 
The LES simulations are performed using several different resolutions smaller than that of the DNS, which is explored in Sec. \ref{sub:generalize} with baseline resolution of $64^3$. 

\begin{figure}
    \centering
    \includegraphics[width=0.38\linewidth]{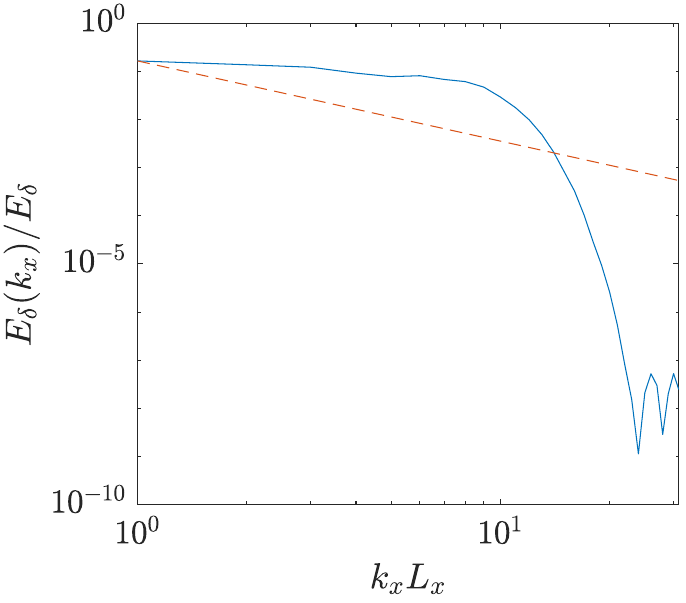} ~ 
    \hspace{-0.37\textwidth}(a)\hspace{0.35\textwidth}
    \includegraphics[width=0.38\linewidth]{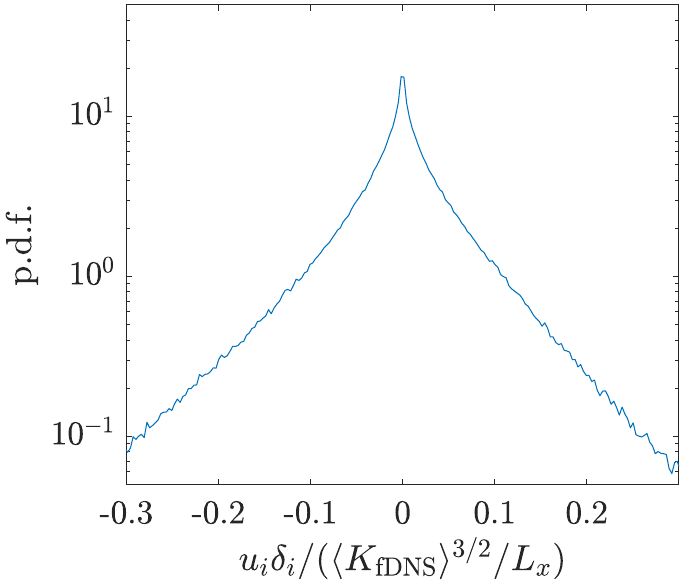}
    \hspace{-0.37\textwidth}(b)\hspace{0.35\textwidth}
    \caption{
    (a) Spectrum of the numerical deviation $u_i\delta_i$ (solid line). Dashed line indicates the -5/3 law. 
    (b) The p.d.f. of the energy transfer for the numerical deviation $u_i\delta_i$ for filter width $\sigma/\Delta_{\textrm{LES}}=2.0$. 
    }
    \label{fig:training-delta}
\end{figure}

{
In Fig. \ref{fig:training-delta}, we show the spectrum for the numerical deviation $\delta_i$ and the p.d.f. of its nondimensional energy transfer $u_i\delta_i$. 
Compared to the -5/3 scaling in the energy spectrum of homogeneous isotropic turbulence, the numerical deviation acts mostly on the intermediate scale. 
In addition, we observe in Fig. \ref{fig:training-delta} (b) that the distribution of the energy transfer $u_i\delta_i$ is symmetric around 0. 
Therefore, the numerical deviation term is intended for modifying the spatial distribution of the energy transfer, instead of changing the global energy transfer amount. 
} 

In the DNS-aided LES framework, all necessary quantities for training a data-driven SGS model are computed. For instance, the LES velocity field is obtained by filtering the DNS velocity field, as illustrated in Fig. \ref{fig:intro-NumErr}, and is used to calculate local invariants as described in Sec. \ref{sub:training}. Additionally, quantities derived from the DNS data that require modeling, specifically, the exact SGS stress $\tau_{ij}$ and the numerical deviation $\delta_i$, are collected as training targets. Once the SGS model is trained on these data, it is applied in an actual LES simulation for \textit{a posteriori} testing. Ideally, these LES simulations should closely replicate the DNS-aided LES results and accurately predict the filtered DNS fields. The generalizability of the trained models is further examined in Sec. \ref{sub:generalize}.

\subsection{Training details}
\label{sub:training}


We use the DNS-aided LES data for training data-driven SGS models. 
To achieve generalizability, we use the model form that follows the physical constraint of Galilean invariance. 
The model involves the strain-rate tensor $\mathbf{S}$ and the rotation-rate tensor $\mathbf{R}$ directly computed from the LES flow field, where
\begin{equation}
    S_{ij} = \frac{1}{2}\left( \frac{\partial \overline{u}_i}{\partial x_j} + \frac{\partial \overline{u}_j}{\partial x_i} \right), ~~~ R_{ij} = \frac{1}{2}\left( \frac{\partial \overline{u}_i}{\partial x_j} - \frac{\partial \overline{u}_j}{\partial x_i} \right). 
    \label{eq:SR}
\end{equation}
The six scalar invariants are computed from $\mathbf{S}$ and $\mathbf{R}$ as 
\begin{equation}
    \begin{array}{lll}
    I_1 = \textrm{tr}({\mathbf{S}}^2), & I_2 = \textrm{tr}({\mathbf{R}}^2), & I_3 = \textrm{tr}({\mathbf{S}}^3),  \\
    I_4 = \textrm{tr}({\mathbf{S}}\,{\mathbf{R}}^2), &
    I_5 = \textrm{tr}({\mathbf{S}}^2{\mathbf{R}}^2), & 
    I_6 = \textrm{tr}({\mathbf{S}}^2{\mathbf{R}}^2{\mathbf{S}}\,{\mathbf{R}}). 
    \end{array}
    \label{eq:inv-scalar}
\end{equation}
At the same time, the four tensor invariants are also linear or nonlinear forms of $\mathbf{S}$ and $\mathbf{R}$ \cite{pope1975more}, 
\begin{equation}
    \begin{array}{ll}
        \mathbf{m}_1=\mathbf{S}, & \mathbf{m}_2=\mathbf{S}^2, \\
        \mathbf{m}_3=\mathbf{R}^2, & \mathbf{m}_4=\mathbf{S} \mathbf{R}-\mathbf{R} \mathbf{S}, \\
    \end{array}
    \label{eq:inv-tensor}
\end{equation}
In the present work, two major model forms are used: the eddy viscosity model 
\begin{equation}
    \boldsymbol{\tau}_{\textrm{ev}} = -2\nu_T(I_1,...,I_6)\mathbf{m_1}, 
    \label{eq:model-1term}
\end{equation} 
and the complex nonlinear model 
\begin{align}
    \boldsymbol{\tau}_{\textrm{cn}} = &-2\nu_{T,1}(I_1,...,I_6)\mathbf{m_1} -2\nu_{T,2}(I_1,...,I_6)\mathbf{m_2} \nonumber\\
&-2\nu_{T,3}(I_1,...,I_6)\mathbf{m_3} -2\nu_{T,4}(I_1,...,I_6)\mathbf{m_4}. 
    \label{eq:model-4terms}
\end{align}
The eddy-viscosity model is dissipative when $\nu_T$ is positive definite and the corresponding dissipation is given by $\varepsilon_{\textrm{SGS}}$. 
The complex nonlinear model includes additional terms to incorporate the anisotropic behavior of the SGS terms.  

To capture the nonlinear behavior between the SGS stress tensor $\boldsymbol{\tau}$ and the scalar invariants, an artificial neural network is trained to find the coefficients in front of the tensor invariants for each model, i.e., the eddy viscosity $\nu_T$ in the eddy-viscosity model and $\nu_{T,1}, \nu_{T,2}, \nu_{T,3}, \nu_{T,4}$ in the complex nonlinear model. 
Both the input and the output features are locally nondimensionalized by the characteristic length scale and the strain-rate scale 
\begin{equation}
    l_{\textrm{char}} = \sqrt{\Delta x^2 + \Delta y^2 + \Delta z^2}, \quad S_{\textrm{char}} = \frac{1}{3}\sqrt{2(\|\mathbf{S}\|^2 + \|\mathbf{R}\|^2)}, 
    \label{eq:char-scales}
\end{equation} 
where the grid size (thus the LES grid truncation scale) in each direction is denoted by $\Delta x, \Delta y, \Delta z$. 
The $1/3$ scaling of the characteristic strain-rate scale is intended to scale in the input features to a similar order of magnitude, which would be essential for a stabilized training \cite{lecun2002efficient}. 
The nondimensionalized quantities are indicated by the superscript $*$. 

We use a fully-connected feedforward neural network with three hidden layers, which have 20 neurons in each layer with the sigmoid function as the activation function. 
Due to the nondimensionalization, the input features $I_1^*+I_2^*=\textrm{const}$, and thus the input layer does not have a bias parameter. 
The hyperparameters are determined through trial and error, and they provide efficient training without the risk of overfitting.

The dataset used for model development is obtained from the DNS-aided LES described in Sec. \ref{sub:DNSaidLES}. 
A portion of the dataset is allocated for training and validation, with a 70:30 split between the two, totaling 1.1e5 points. 
A separate subset is reserved for testing, which includes both \textit{a priori} and \textit{a posteriori} evaluations, as discussed in Sec. \ref{sec:results}. 
Model training is conducted using the Adam optimization algorithm \cite{kingma2014adam}, with an initial learning rate of 0.01. 
The learning rate is adaptively reduced if the validation error stagnates. Training is terminated when the loss function plateaus, which typically occurs after approximately 100 epochs. 
{
The training history of the models is shown in the appendix.} 

The loss function is given by the $\mathcal{L}_2$ loss of the local dissipation rate. 
{
For comparison, both a model excluding the numerical deviation and a model including the numerical deviation are considered during training; the model formulation and the model implementation are identical between the two models.} 
For a model excluding numerical deviation, the loss function is given by
\begin{equation}    
    L=\mathcal{L}_2(\varepsilon^*, \varepsilon_{\textrm{model}}^*)=\sum (S_{ij}^*\tau_{ij}^* - S_{ij}^*\tau_{\textrm{model},ij}^*)^2,
    \label{eq:loss-nodelta}
\end{equation}
{
where $\varepsilon^* = \varepsilon_{\textrm{SGS}}^* = - S_{ij}^*\tau_{ij}^*$ is commonly used in literature for model evaluation \cite{lilly1992proposed,zhou2019subgrid}.} 
For a model including the numerical deviation, the loss function is given by
\begin{align}
    L=\mathcal{L}_2(\varepsilon^*,\varepsilon_{\textrm{model}}^*) & = \sum\left[\left(u_i^*\frac{\partial \tau_{ij}^*}{\partial x_j^*} - u_i^*\delta_i^*\right) - u_i^*\frac{\partial \tau_{\textrm{model},ij}^*}{\partial x_j^*}\right]^2, \\
    \textrm{where}~\varepsilon^* & = -\mathcal{D}_{\textrm{SGS}}^*+\varepsilon_{\textrm{SGS}}^*+\varepsilon_{\textrm{numerical}}^* = u_i^*\frac{\partial \tau_{ij}^*}{\partial x_j^*} - u_i^*\delta_i^*. 
    \label{eq:loss-delta}
\end{align}
Here, the subscript ``model'' indicates a modeled quantity for both the eddy-viscosity and complex nonlinear models. 
{
The dissipation rate including the numerical deviation is turned into the vector form since the numerical deviation is naturally a vector in Eqn. \eqref{eq:NS-delta-momonly}.} 
The superscript $*$ indicates that the quantities are nondimensionalized using characteristic scales given by Eq. \eqref{eq:char-scales}. 
Note that the loss function is designed to minimize the error in the local dissipation rate using the data-driven methods instead of the domain-averaged dissipation rate in the non-data-driven methods. 
As observed in the previous literature \cite{lund1993parameterization}, a constant eddy viscosity does not necessarily provide the optimal local dissipation rate, and the local values have strong scattering. 

\section{Impact of the numerical deviation}
\label{sec:results}


In this section, we examine the impact of the numerical deviation in both \textit{a priori} and \textit{a posteriori} tests. 
The training performance is evaluated and compared between models excluding and including the numerical deviation. 
All the results are based on the baseline filter width $\sigma/\Delta_{\textrm{LES}}=2.0$. 

\subsection{\textit{A priori} training performance}
\label{sub:apriori}




\begin{figure}
    \centering
    \includegraphics[width=0.28\linewidth, trim={0 0 1.5cm 0}, clip]{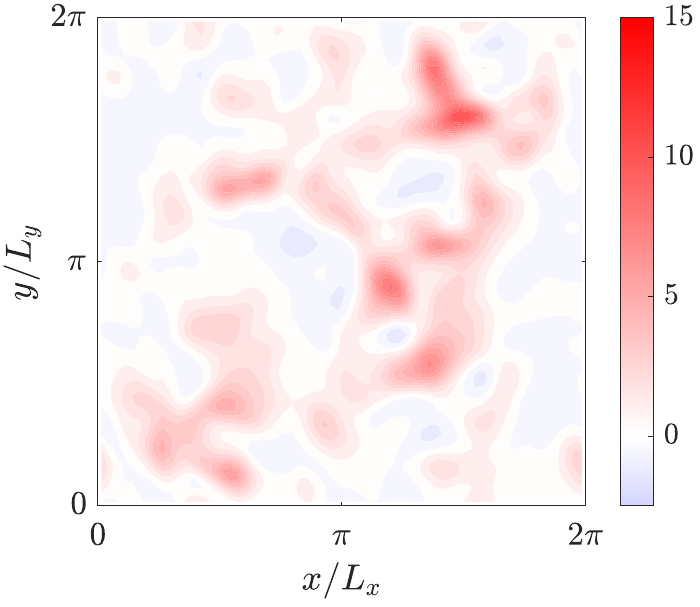} 
    \hspace{-0.27\textwidth}(a)\hspace{0.25\textwidth}
    \includegraphics[width=0.30\linewidth, trim={0.8cm 0 0 0}, clip]{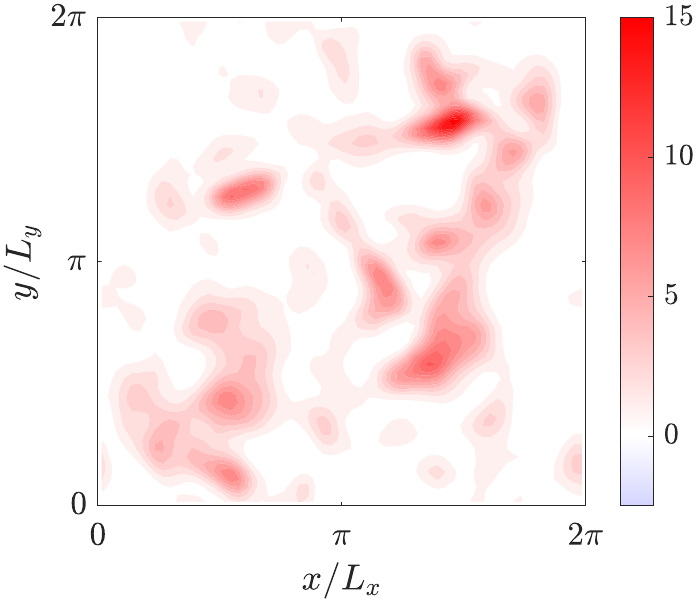} 
    \hspace{-0.29\textwidth}(b)\hspace{0.27\textwidth}
    \includegraphics[width=0.30\linewidth]{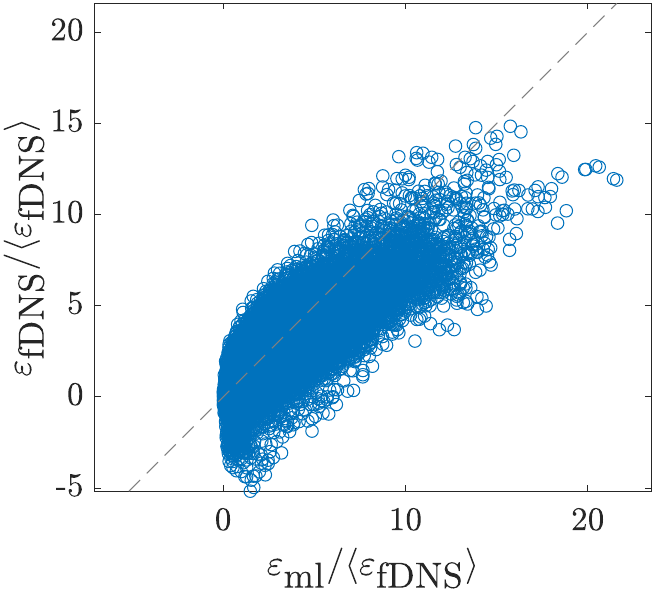} 
    \hspace{-0.29\textwidth}(c)\hspace{0.27\textwidth} \phantom{,}
    \\
    \vspace{3mm}
    \includegraphics[width=0.28\linewidth, trim={0 0 1.5cm 0}, clip]{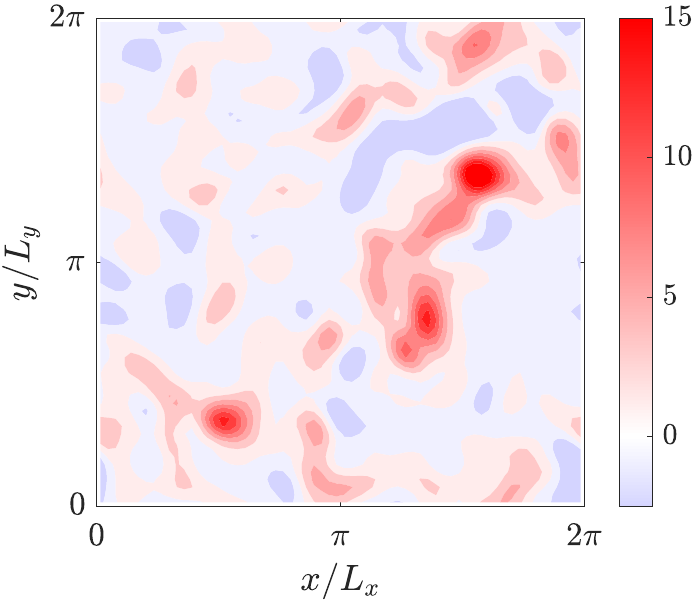} 
    \hspace{-0.27\textwidth}(d)\hspace{0.25\textwidth}
    \includegraphics[width=0.30\linewidth, trim={0.8cm 0 0 0}, clip]{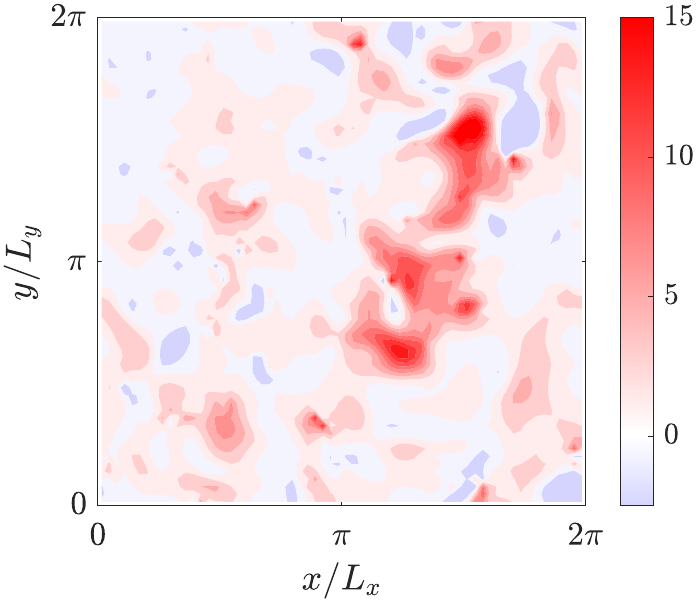} 
    \hspace{-0.29\textwidth}(e)\hspace{0.27\textwidth}
    \includegraphics[width=0.30\linewidth]{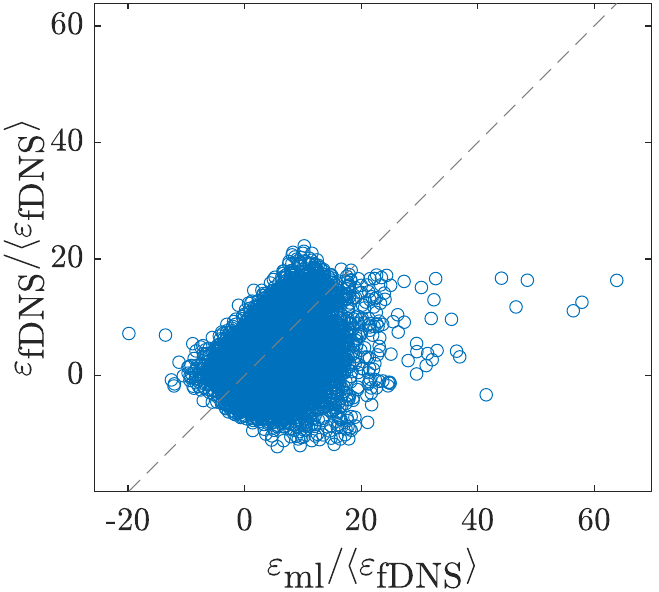} 
    \hspace{-0.29\textwidth}(f)\hspace{0.27\textwidth} \phantom{,} 
    \caption{Instantaneous (a,d) target dissipation rate $\varepsilon/\langle \varepsilon_{\textrm{fDNS}}\rangle$ and 
    (b,e) modeled dissipation rate $\varepsilon_{\textrm{ml}}/\langle \varepsilon_{\textrm{fDNS}}\rangle$ on the same $x-y$ plane and
    (c,f) scatter plot showing the correlation between $\varepsilon_{\textrm{ml}}$ and $\varepsilon_{\textrm{fDNS}}$ 
    for eddy-viscosity neural-network SGS models (a,b,c) excluding and (d,e,f) including the numerical deviation for baseline filter width $\sigma/\Delta_{\textrm{LES}}=2.0$. 
    The dashed line in (c,f) indicates identity. }
    \label{fig:apriori-1term}
\end{figure}

The trained models are first evaluated in the \textit{a priori} tests, using the filtered DNS data. 
For each evaluation, we compare the trained models excluding the numerical deviation and including the numerical deviation. 
In order to visualize the turbulent structures of different scales, the nondimensionalization is based on the global dissipation of the filtered DNS $\langle \varepsilon_{\textrm{fDNS}}\rangle$, where $\langle \cdot \rangle$ indicates the spatial averaging. 

\begin{table}[]
    \centering
    \begin{tabular}{c|c|c|c|c|c}
        \hline \hline 
         &  & EV-noND & ~~EV-ND~~ & CN-noND & ~~CN-ND~~ \\ \hline
        \multirow{2}{*}{$\sigma/\Delta_{\textrm{LES}}=0.5$} 
        & rRMSE & 0.7094 & 1.0149 & 0.7711 & 0.9511 \\ \cline{2-6} 
        & Corr. & 0.6366 & 0.2456 & 0.6829 & 0.3860 \\ 
        \hline
        \multirow{2}{*}{$\sigma/\Delta_{\textrm{LES}}=1.0$} 
        & rRMSE & 0.5021 & 0.9538 & 0.7417 & 0.6701 \\ \cline{2-6} 
        & Corr. & 0.8322 & 0.4026 & 0.8079 & 0.7777 \\ 
        \hline
        {$\sigma/\Delta_{\textrm{LES}}=2.0$} 
                   & rRMSE & 0.4947 & 1.0524 & 0.7624 & 0.7543 \\ \cline{2-6} 
        (Baseline) & Corr. & 0.8764 & 0.4181 & 0.7384 & 0.7952 \\ 
        \hline
        \multirow{2}{*}{$\sigma/\Delta_{\textrm{LES}}=4.0$} 
        & rRMSE & 0.3850 & 0.7929 & 1.4408 & 0.7449 \\ \cline{2-6} 
        & Corr. & 0.8760 & 0.5092 & 0.4555 & 0.7370 \\ 
        \hline \hline
    \end{tabular}
    \caption{\textit{A priori} evaluation of the root mean square errors (rRMSE) and Pearson correlation coefficient (Corr.) of the dissipation rate $\varepsilon/\langle \varepsilon_{\textrm{fDNS}}\rangle$.
    EV indicates eddy-viscosity model, and CN indicates complex nonlinear model. 
    The models excluding and including the numerical deviation are indicated by noND and ND, respectively. }
    \label{tab:apriori}
\end{table}

Figure \ref{fig:apriori-1term} presents the \textit{a priori} evaluation of the trained eddy-viscosity-type SGS models, comparing the versions that exclude and include the numerical deviation. 
{
To qualitatively assess model accuracy across different scales, we compare the target dissipation rate with that predicted by the trained models as given by the loss function. 
Note that the two models have different loss functions, and the target dissipation rate is not identical between the two.} 
The correlation between the modeled and target dissipation rates is shown in the form of scatter plots. 
Quantitative evaluation is provided in Table \ref{tab:apriori}, which reports both the Pearson correlation coefficient and the relative error rate. 
It is important to note that, due to differences in characteristic scales, this \textit{a priori} evaluation differs from the metrics used during model training.

When the numerical deviation is excluded, the target dissipation rate shown in Fig. \ref{fig:apriori-1term}(a) is predominantly positive, reflecting a locally dissipative behavior. As shown in Fig. \ref{fig:apriori-1term}(b), the eddy-viscosity model captures the large-scale features of the dissipation field reasonably well, but fails to accurately reproduce the finer-scale structures. The scatter plot in Fig. \ref{fig:apriori-1term}(c) indicates a strong correlation between the predicted and true dissipation values, supported by a high Pearson correlation coefficient.

In comparison, when the numerical deviation is included, the target dissipation rate in Fig. \ref{fig:apriori-1term}(d) is no longer strictly locally dissipative, and exhibits stronger spatial variability. As shown in Fig. \ref{fig:apriori-1term}(e), the eddy-viscosity neural network model captures a broader range of flow structures, but fails to accurately reproduce both the large-scale and small-scale features. The scatter plot in Fig. \ref{fig:apriori-1term}(f) confirms this, showing weaker correlation between the predicted and true values. The model occasionally exhibits large deviations, resulting in higher error in the \textit{a priori} evaluation. This outcome is expected, as including the numerical deviation introduces additional complexity, making it more difficult to extract accurate models when relying solely on local quantities.

\begin{figure}
    \centering
    \includegraphics[width=0.28\linewidth, trim={0 0 1.5cm 0}, clip]{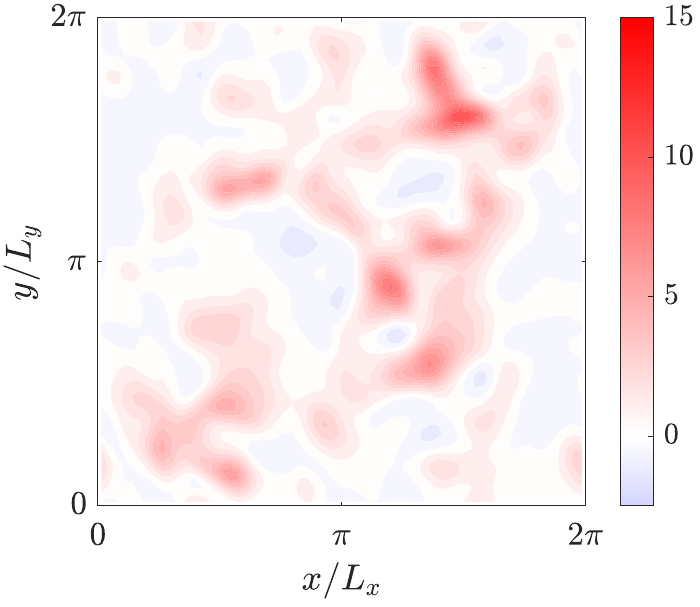} 
    \hspace{-0.27\textwidth}(a)\hspace{0.25\textwidth}
    \includegraphics[width=0.30\linewidth, trim={0.8cm 0 0 0}, clip]{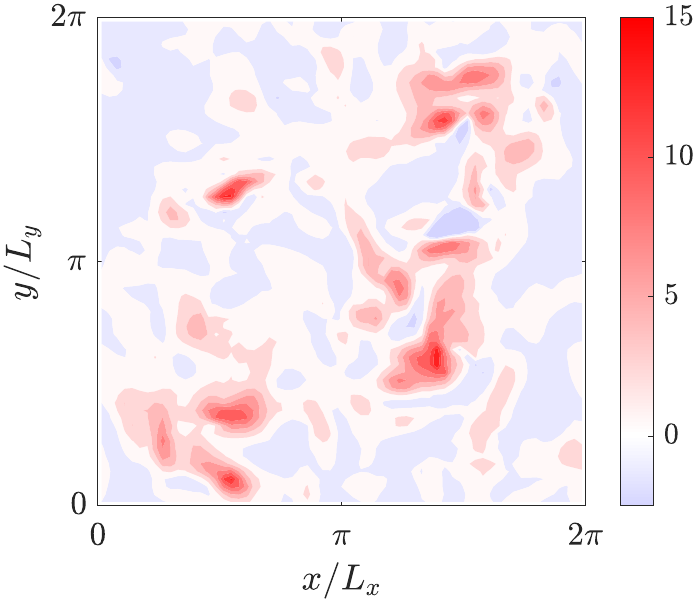} 
    \hspace{-0.29\textwidth}(b)\hspace{0.27\textwidth}
    \includegraphics[width=0.31\linewidth]{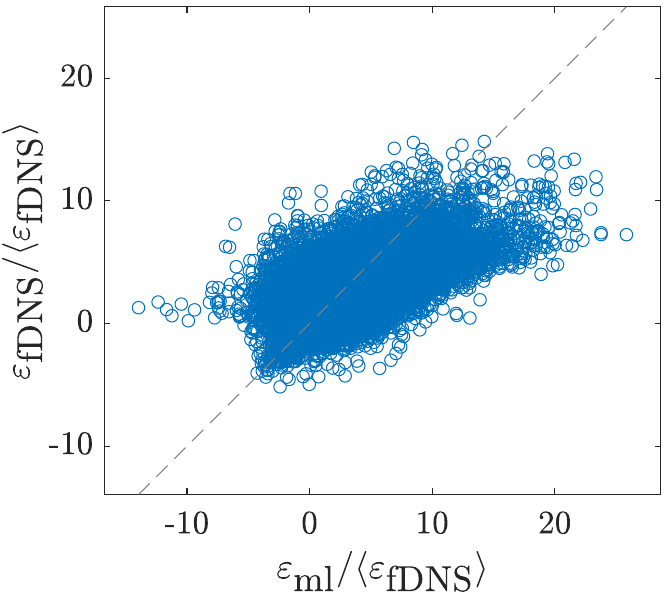} 
    \hspace{-0.30\textwidth}(c)\hspace{0.28\textwidth} \phantom{,}
    \\
    \vspace{3mm}
    \includegraphics[width=0.28\linewidth, trim={0 0 1.5cm 0}, clip]{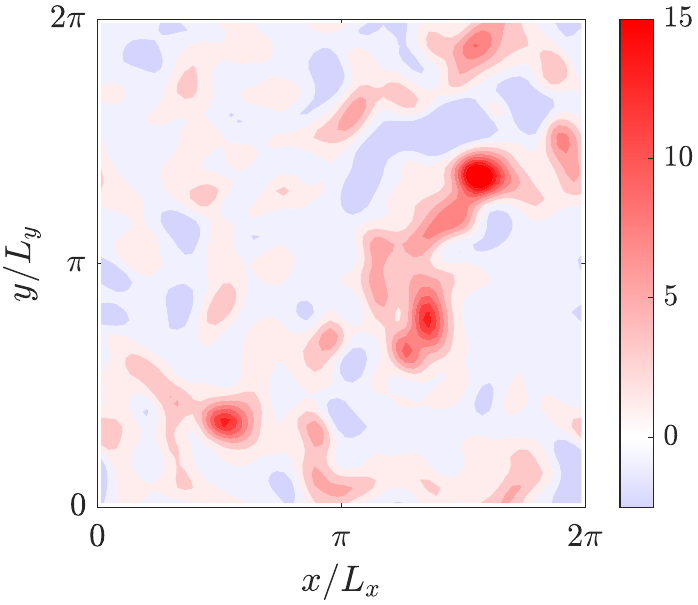} 
    \hspace{-0.27\textwidth}(d)\hspace{0.25\textwidth}
    \includegraphics[width=0.30\linewidth, trim={0.8cm 0 0 0}, clip]{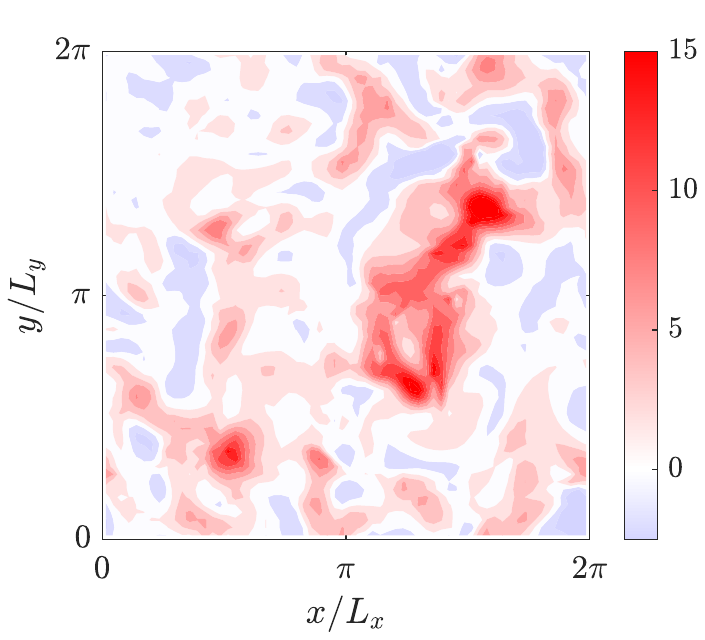} 
    \hspace{-0.29\textwidth}(e)\hspace{0.27\textwidth}
    \includegraphics[width=0.31\linewidth]{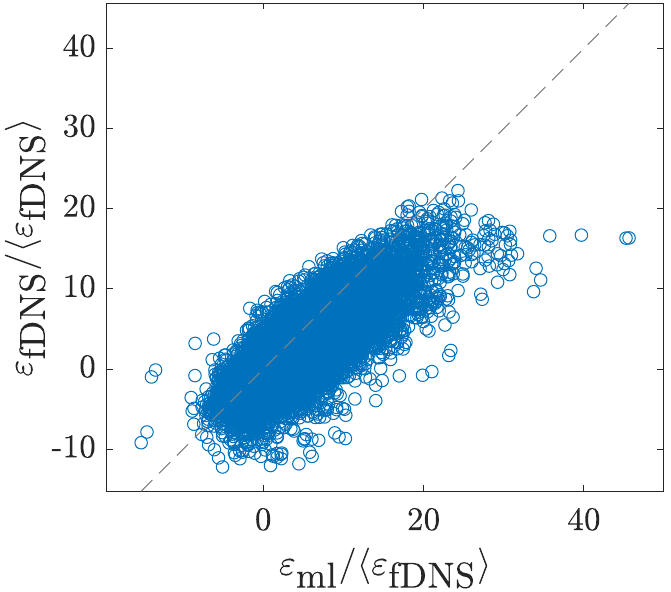} 
    \hspace{-0.30\textwidth}(f)\hspace{0.28\textwidth} \phantom{,}
    \caption{Instantaneous (a,d) target dissipation rate $\varepsilon/\langle \varepsilon_{\textrm{fDNS}}\rangle$ and 
    (b,e) modeled dissipation rate $\varepsilon_{\textrm{ml}}/\langle \varepsilon_{\textrm{fDNS}}\rangle$ on the same $x-y$ plane and
    (c,f) scatter plot showing the correlation between $\varepsilon_{\textrm{ml}}$ and $\varepsilon_{\textrm{fDNS}}$ 
    for complex nonlinear neural-network SGS models (a,b,c) excluding and (d,e,f) including the numerical deviation for baseline filter width $\sigma/\Delta_{\textrm{LES}}=2.0$. 
    The dashed line in (c,f) indicates identity. 
    }
    \label{fig:apriori-4term}
\end{figure}


In Fig. \ref{fig:apriori-4term}, the visualization of the complex nonlinear model’s performance, given by Eq. \eqref{eq:model-4terms}, reveals markedly different behavior compared to the eddy-viscosity model. When the numerical deviation is excluded, as shown in Fig. \ref{fig:apriori-4term}(a,b), the model prediction exhibits rich small-scale structures, benefiting from the increased expressiveness of the model form. However, the large-scale structures are not as well captured as in the eddy-viscosity case, as they are overshadowed by the amplified small-scale activity. Furthermore, the model is no longer locally dissipative, a consequence of the more flexible nonlinear formulation. The impact of this increased complexity is discussed further in Sec. \ref{sub:apost}. Despite these differences, the model still achieves a high correlation between predicted and true dissipation values, as seen in Fig. \ref{fig:apriori-4term}(c), although local dissipation is often overestimated.

When the numerical deviation is included (Fig. \ref{fig:apriori-4term}(d,e)), the model more accurately captures both large-scale and small-scale structures. Notably, the correlation between predictions and true values, shown in Fig. \ref{fig:apriori-4term}(f), is significantly improved compared to the eddy-viscosity model. This suggests that the complex nonlinear model, when trained with the numerical deviation, can better reconcile multiple scales of turbulence.


\subsection{\textit{A posteriori} results}
\label{sub:apost}



\begin{figure}
    \centering
    \includegraphics[width=0.6\linewidth]{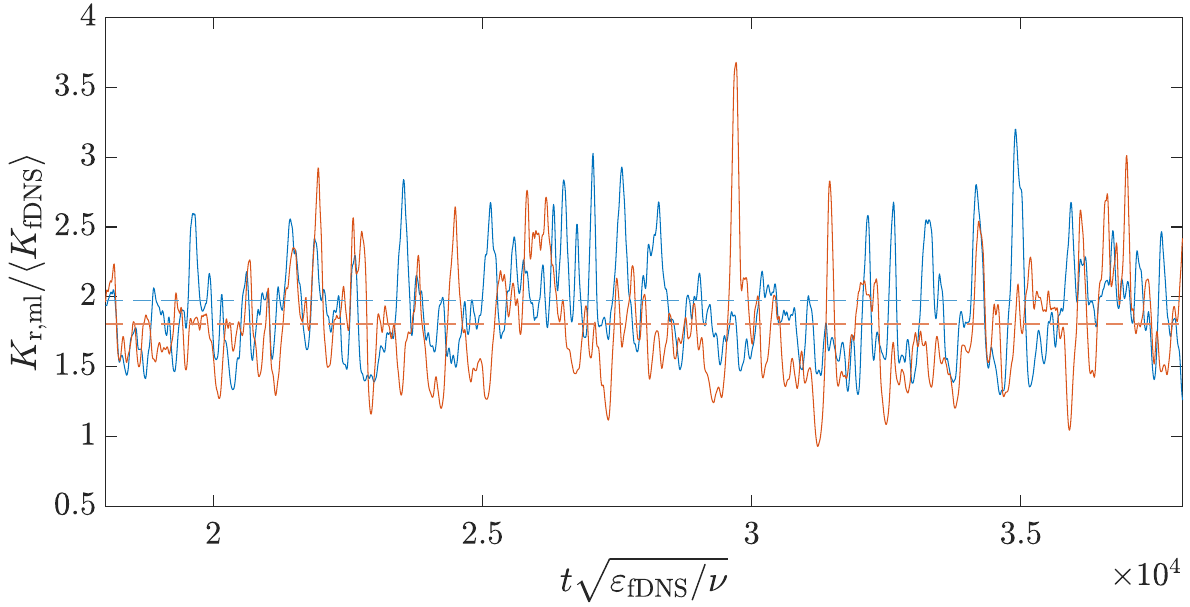} ~
    \hspace{-0.59\textwidth}(a)\hspace{0.57\textwidth}
    \includegraphics[width=0.35\linewidth]{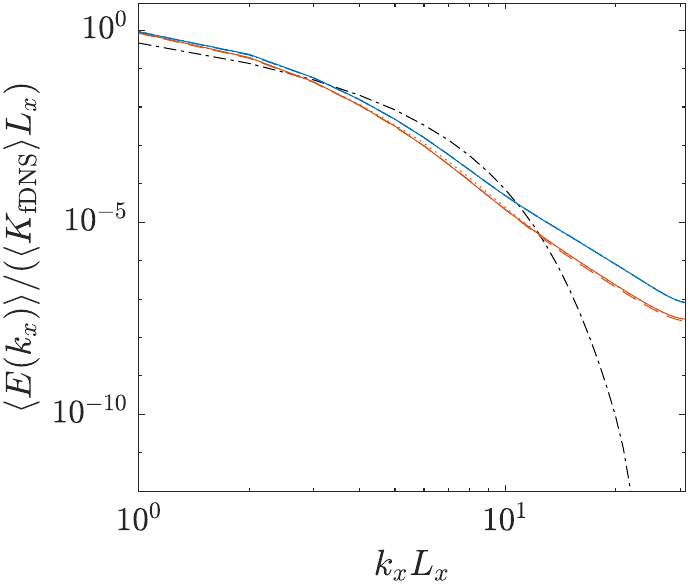}
    \hspace{-0.34\textwidth}(b)\hspace{0.32\textwidth}
    \caption{
    (a) Time evolution of the instantaneous (solid) and average (dashed) kinetic energy for eddy-viscosity SGS model excluding (blue) and including (red) the numerical deviation. 
    (b) Energy spectra of the \textit{a posteriori} for eddy-viscosity SGS model excluding (blue) and including (red) the numerical deviation using time step of $\Delta t_{\textrm{DNS}}$ (solid), $2\Delta t_{\textrm{DNS}}$ (dashed), and $5\Delta t_{\textrm{DNS}}$ (dotted).
    Energy spectrum of the filtered DNS is given by the black dot-dashed line.}
    \label{fig:apost-1term}
\end{figure}

\begin{figure}
    \centering
    \includegraphics[width=0.6\linewidth]{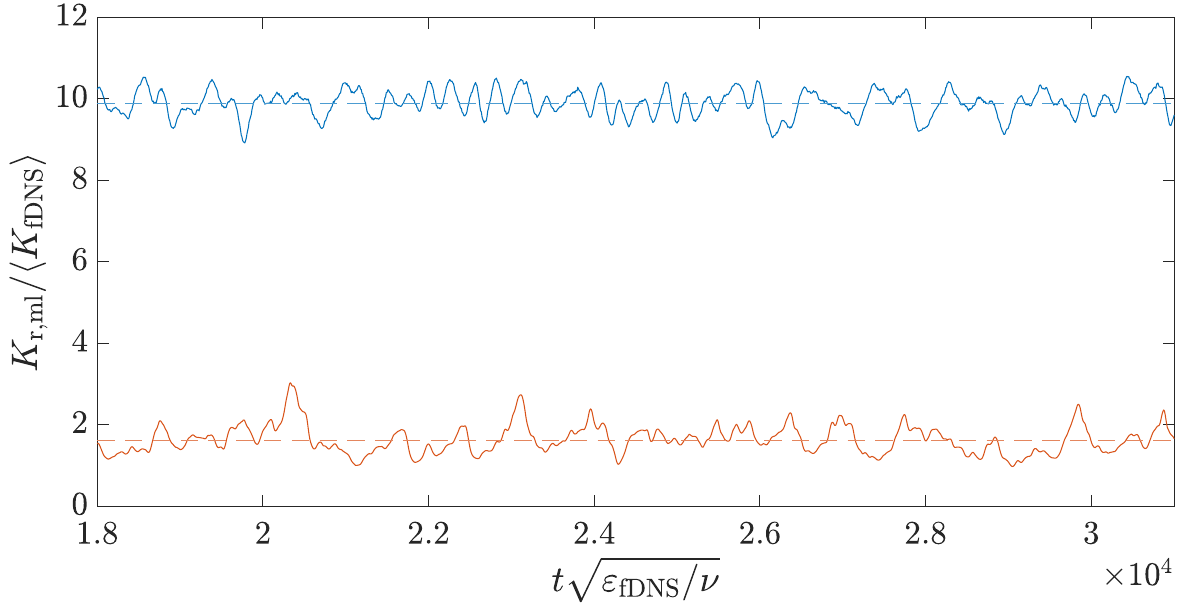} ~
    \hspace{-0.59\textwidth}(a)\hspace{0.57\textwidth}
    \includegraphics[width=0.35\linewidth]{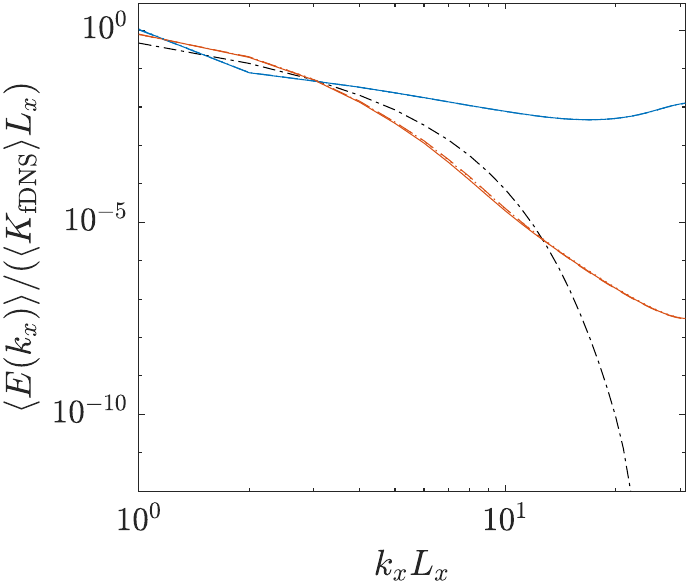}
    \hspace{-0.34\textwidth}(b)\hspace{0.32\textwidth}
    \caption{(a) Time evolution of the instantaneous (solid) and average (dashed) kinetic energy for complex nonlinear SGS model excluding (blue) and including (red) the numerical deviation. 
    (b) Energy spectra of the \textit{a posteriori} for complex nonlinear SGS model excluding (blue) and including (red) the numerical deviation using time step of $\Delta t_{\textrm{DNS}}$ (solid), $2\Delta t_{\textrm{DNS}}$ (dashed), and  $5\Delta t_{\textrm{DNS}}$ (dotted).
    Energy spectrum of the filtered DNS is given by the black dot-dashed line.}
    \label{fig:apost-4terms}
\end{figure}

In the data-driven methods, the \textit{a posteriori} tests often fail to align with the \textit{a priori} evaluations. 
To improve this consistency, we incorporate the numerical deviation in model training. 
We assess the resulting consistency by examining the energy spectra from \textit{a posteriori} simulations using models trained with and without the numerical deviation. 
For the \textit{a posteriori} tests, the trained neural-network models are implemented within the second-order finite-difference LES code described in Sec. \ref{sub:DNSaidLES}. 
The filtered DNS flow fields serve as the initial conditions for the LES. 
To ensure a fair comparison, the simulations are run using the same time step as was used to generate the training data. However, it is important to note that this choice is not a constraint of the modeling framework; the flexibility in time stepping is discussed further in Sec. \ref{sub:generalize}.
The simulations are run until the kinetic energy reaches a statistically converged state as demonstrated in Fig. \ref{fig:apost-1term}(a) and Fig. \ref{fig:apost-4terms}(a). 
As a reference, the kinetic energy of the filtered DNS fluctuates $\pm50\%$ around unity value though not shown here, consistent with the fluctuation seen in the literature \cite{lundgren2003linearly}. 
No additional stabilization strategies, such as backscatter clipping, are applied during the \textit{a posteriori} simulations.

In Fig. \ref{fig:apost-1term}, we evaluate the \textit{a posteriori} performance of the eddy-viscosity neural-network SGS models trained both with and without the numerical deviation. 
Fig. \ref{fig:apost-1term}(a) shows the time evolution of the kinetic energy from these simulations. 
Both simulations remain numerically stable and evolve toward statistically steady states with higher kinetic energy than the filtered DNS. 
{
Notably, the model trained with the numerical deviation yields slightly lower kinetic energy, indicating more dissipative behavior. 
Though both models are trained for accurate kinetic energy transfer, excluding the numerical deviation will neglect the possible impact of the numerical deviation on the energy spectrum. 
This leads to different \textit{a posteriori} behavior of the kinetic energy transfer. 
The interaction between the model and the LES solver will further exaggerate such differences.} 

This observation is further supported by the energy spectra in Fig. \ref{fig:apost-1term}(b). Both models show reasonable agreement with the filtered DNS across a range of scales, although the small-scale turbulence is generally overpredicted. This overprediction is consistent with the \textit{a priori} results, where neither model accurately captures the fine-scale structure of the local dissipation rate. Interestingly, both models tend to underpredict intermediate-scale turbulence, though the precise mechanism remains unclear.

Comparing the two, the inclusion of the numerical deviation leads to a modest suppression of small-scale energy, which contributes to slightly better spectral alignment with the filtered DNS. Although the model trained with the numerical deviation performs worse in terms of \textit{a priori} correlation and error metrics, it achieves comparable—if not slightly improved—\textit{a posteriori} performance. This result highlights a key insight: minimizing the \textit{a priori} error in the SGS dissipation rate does not necessarily ensure better \textit{a posteriori} performance. Numerical effects, particularly the numerical deviation, play a critical role and must be accounted for during model development.

In Fig. \ref{fig:apost-4terms}, we present the \textit{a posteriori} evaluation of the complex nonlinear SGS models trained with and without the inclusion of numerical deviation. The use of a richer model form with additional terms significantly enhances the representation of small-scale turbulence in both \textit{a priori} and \textit{a posteriori} tests. However, this added complexity also increases the risk of numerical instability, primarily due to the higher likelihood of local energy backscattering. To maintain a fair evaluation, no backscatter clipping or artificial stabilization is applied.

As shown in Fig. \ref{fig:apost-4terms}(a), the model trained without accounting for numerical deviation results in an \textit{a posteriori} simulation where the kinetic energy grows well beyond the expected level, even after convergence. The corresponding energy spectrum in Fig. \ref{fig:apost-4terms}(b) confirms that the small-scale turbulence is severely overpredicted, and the inaccuracy extends across all resolved scales, distorting the entire spectral shape.

In contrast, the model trained with the numerical deviation yields a \textit{a posteriori} spectrum that closely aligns with the filtered DNS. While the small-scale turbulence remains slightly overpredicted, the magnitude of the deviation is within an acceptable range, and the overall spectrum is much more physical. Furthermore, this model demonstrates improved numerical stability throughout the simulation.

These results emphasize the crucial role of including numerical deviation in training, especially for high-capacity models. Accounting for the numerical discrepancy helps constrain the model in regimes where nonlinear effects dominate, leading to improved physical realism and stable LES performance.

In the \textit{a priori} tests, the neural network models that exclude the numerical deviation perform well for both the eddy-viscosity and the complex nonlinear model forms. 
However, these promising \textit{a priori} results do not translate to consistent \textit{a posteriori} performance. 
In particular, the complex nonlinear model trained without accounting for numerical deviation fails to yield meaningful LES simulations. 
{
Despite careful model design and nondimensionalization grounded in physical principles, the combination of model complexity and strong neural network nonlinearity could bring the \textit{a posteriori} solutions beyond available training data. 
Therefore, a data-driven model lacking generalizability could become numerically unstable, especially for a complex model.} 

For homogeneous isotropic turbulence, a simpler eddy-viscosity-type model is often sufficient. Yet, when modeling more complex flows—where anisotropy, non-equilibrium effects, or additional physical mechanisms are present—more expressive models with nonlinear terms become necessary to capture the essential small-scale dynamics. Unfortunately, the added model capacity also amplifies the inconsistency between the \textit{a priori} and \textit{a posteriori} tests, making data-driven SGS model development for such flows particularly challenging.

To mitigate this inconsistency, incorporating the numerical deviation during training proves to be an effective strategy. This approach helps align the behavior of the model in \textit{a posteriori} simulations with its \textit{a priori} evaluations, thereby enhancing the physical relevance of \textit{a priori} tests and improving the reliability of the training process.


\section{Discussion}
\label{sec:discuss}


\subsection{The relationship to the non-data-driven models}
\label{sub:const-results}



Numerical deviation is always present, regardless of whether data-driven techniques are used for model development. 
However, data-driven models tend to exhibit a more pronounced inconsistency between the training process and their actual application in LES simulations. 
To better understand the impact of data-driven methods on \textit{a posteriori} performance, {
we carry out tests} that bridge the gap between data-driven and traditional (non-data-driven) models. 
Specifically, instead of using neural networks to predict the coefficients of tensor invariants in the eddy-viscosity and complex nonlinear models, we directly ``train'' constant coefficients from the data. 
This approach is equivalent to applying linear regression on the local tensor invariants $\mathbf{m}_1, \mathbf{m}_2, \mathbf{m}_3, \mathbf{m}_4$ and allows us to isolate the effects of model nonlinearity from the effects of data-driven optimization.
Note that the eddy-viscosity model with a constant coefficient is similar to the traditional Smagorinsky model, 
\begin{equation}
    \boldsymbol{\tau}_{\textrm{Smag}} = -2(C_s\Delta)^2|\mathbf{S}|\mathbf{S} = - 2(C_s\Delta)^2|\mathbf{S}|\mathbf{m}_1 , 
    \label{eq:model-Smag}
\end{equation}
where $C_s=0.16$ is the typical Smagorinsky coefficient \cite{lilly1992proposed}, and $\Delta$ is the resolved length scale. 
Similarly, the complex nonlinear model with constant coefficients is similar to the mixed model, which combines the tensor diffusivity model with the eddy-viscosity term \cite{zang1993dynamic}. 
For this model, the SGS stress term is given by
\begin{equation}
    {\tau}_{\textrm{mixed},ij} = \Delta^2\frac{\partial u_i}{\partial x_k}\frac{\partial u_j}{\partial x_k} - 2(C_{s'}\Delta)^2|\mathbf{S}|S_{ij}, 
    \label{eq:model-mixed}
\end{equation}
or equivalently
\begin{equation}
     \boldsymbol{\tau}_{\textrm{mixed}} = - 2(C_{s'}\Delta)^2|\mathbf{S}|\mathbf{m}_1 + \Delta^2(\mathbf{m}_2-\mathbf{m}_3-\mathbf{m}_4),
     \label{eq:model-mixed-2}
\end{equation}
where $C_{s'}$ is the modified Smagorinsky coefficient. Note that the second part of the mixed model is similar to the tensor diffusivity model, which is derived from a Taylor series of the filter operator \cite{stolz2001approximate}. Thus, the mixed model can be seen as the tensor diffusivity model with additional dissipation.

\begin{table}[]
    \centering
    \begin{tabular}{c||c|c|c||c|c|c}
        \hline \hline
         & EV-noND & ~~EV-ND~~ & EV-theory & CN-noND & ~~CN-ND~~ & TD-theory \\
         & -const  & -const  & & -const  & -const   & \\
        \hline
        ~~$\nu_{T,1}^*$~~ & 0.9780 & 1.3323 & 1.2288 & 0.8512 & 1.2519 & ---    \\ 
        \hline
        $\nu_{T,2}^*$     & ---    & ---    & ---    &-0.3046 &-0.4042 &-0.6667 \\ 
        \hline
        $\nu_{T,3}^*$     & ---    & ---    & ---    & 0.8658 & 0.9194 & 0.6667 \\ 
        \hline
        $\nu_{T,4}^*$     & ---    & ---    & ---    & 4.8161 & 0.7701 & 0.6667 \\ 
        \hline \hline
    \end{tabular}
    \caption{Fitted coefficients for eddy-viscosity (EV) model (Eq. \eqref{eq:model-1term}) and complex nonlinear (CN) model (Eq. \eqref{eq:model-4terms}), along with the theoretical prediction of the Smagorinksy model and the tensor diffusivity model. }
    \label{tab:const-coeff}
\end{table}

\begin{figure}
    \centering
    \includegraphics[width=0.56\linewidth]{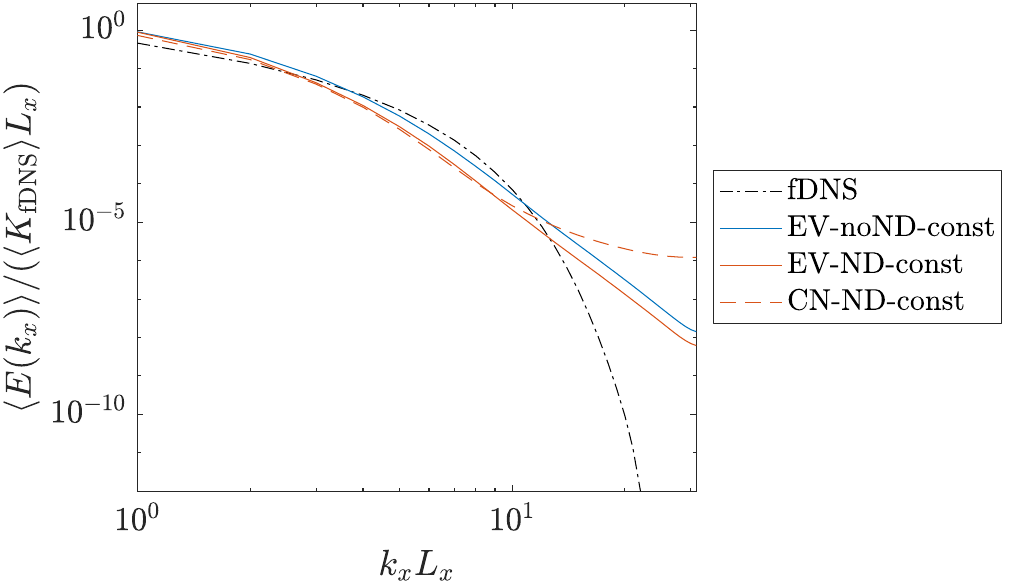}
    \caption{Energy spectra of \textit{a posteriori} analysis of data-driven constant coefficient eddy-viscosity (solid) and complex nonlinear (dashed) models, excluding (blue) and including (red) numerical deviation compared to filtered DNS (black dot-dashed). 
    {
    The \textit{a posteriori} analysis for CN-noND-const is not available due to numerical instability. }
    }
    \label{fig:apost-const}
\end{figure}

The training procedure follows the method outlined in Sec. \ref{sub:training}, with the exception that a constant is used instead of the neural network, resulting in significantly faster training. The resulting constant coefficients are summarized in Table \ref{tab:const-coeff}. For reference, the eddy viscosity of the classical Smagorinsky model is $\nu_T^* = 1.2288$. Among the eddy-viscosity-type models, the data-driven model trained without including the numerical deviation (EV-noND-const) produces a smaller coefficient than expected, whereas the model trained with the numerical deviation (EV-ND-const) yields a larger coefficient. The \textit{a posteriori} results for both models, shown in Fig. \ref{fig:apost-const}, confirm that incorporating the numerical deviation results in a modest suppression of small-scale turbulence—consistent with the trends observed in the neural-network-based models.

On the other hand, the complex nonlinear model bears resemblance to the classical mixed model, which combines an eddy-diffusivity term with the tensor diffusivity model. The tensor diffusivity term for the mixed model, which is derived from the Taylor series of the filter operator \cite{stolz2001approximate}, is not necessarily numerically stable due to lack of dissipation \cite{winckelmans2001explicit}.
In the classical formulation, the theoretical coefficients for the tensor diffusivity terms are given by $\nu_{T,2}^* = -\nu_{T,3}^* = -\nu_{T,4}^* = -2/3$, which we adopt as a reference for evaluating the correspondig coefficients in the data-driven model listed in Table \ref{tab:const-coeff}.

We observe that the trained complex nonlinear model with constant coefficients (CN-noND-const), when excluding the numerical deviation, deviates significantly from the tensor diffusivity component of the mixed model—particularly in the value of $\nu_{T,4}^*$. 
In contrast, the model trained with the numerical deviation included (CN-ND-const) yields coefficients that align more closely with the theoretical values of the tensor diffusivity model, where $\nu_{T,4}^*$ is of the same order of magnitude as $\nu_{T,2}^*$ and $\nu_{T,3}^*$. 
This discrepancy arises because {
$\mathbf{m}_4$}, when evaluated on a staggered grid, appears nearly nondissipative if the SGS dissipation rate is computed solely as $S{ij}^* \tau_{ij}^*$. 
As a result, the model trained without accounting for numerical effects tends to overemphasize the contribution of $\mathbf{m}_4$.

However, $\mathbf{m}_4$ still affects the evolution of local kinetic energy through the SGS transport term and numerical deviation. 
A more appropriate evaluation function in this context is $u_i^*(\partial \tau_{ij}^*/\partial x_j^* - \delta_i^*)$, which captures both physical and numerical contributions. 
As shown in Fig. \ref{fig:apost-const}, the \textit{a posteriori} test for CN-noND-const is not available due to numerical instability. 
Even augmenting the model with an eddy-viscosity term, as in the traditional mixed model, does not stabilize it—likely due to the excessive value of $\nu_{T,4}^*$. 
In contrast, the CN-ND-const model is numerically stable and yields a good match with the filtered DNS energy spectrum.

Furthermore, comparing the neural-network-based CN-ND model in Fig. \ref{fig:apost-4terms}(b) with the CN-ND-const model in Fig. \ref{fig:apost-const} shows that the neural network improves small-scale predictions. 
The additional flexibility from the nonlinear parameterization helps mitigate small-scale overprediction that can arise from complex model structures.

This comparison between the data-driven models and the non-data-driven models indicates the difficulties the data-driven models may have in realizing the physical constraints learned through theoretical derivation, and these difficulties leads to issues such as overpredicting specific nonlinear terms. 
During the \textit{a priori} tests, the impact of the overprediction is not easy to recognize. 
However, as the simulation evolves, the error in the solution accumulates and deviates from the filtered DNS. 
The \textit{a posteriori} tests may have significantly distorted small-scale turbulence and thus could be numerically unstable. 
The data-driven techniques further magnify the issue because the model form and the neural network architecture lead to more nonlinearity in the model. 
The coupling nonlinearity complicates the inconsistency issue and makes it more prominent and more challenging in data-driven modeling. 
In our experiments, including the numerical deviation eases the challenge of such inconsistency between the \textit{a priori} tests and the \textit{a posteriori} tests in data-driven modeling. 

\subsection{Model generalizability} 
\label{sub:generalize}


\begin{figure}
    \centering
    \includegraphics[width=0.38\linewidth, trim={0 0 3.8cm 0}, clip]{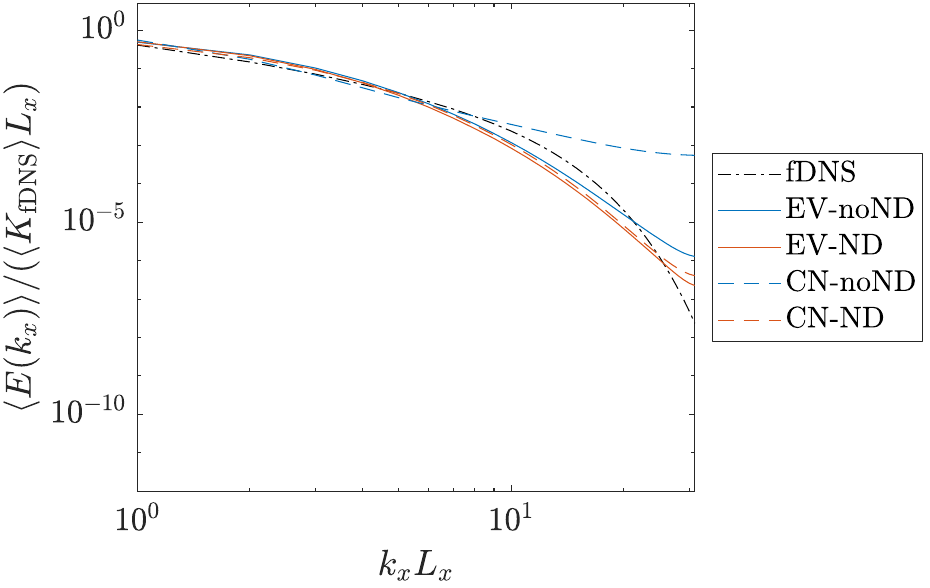} ~ 
    \hspace{-0.37\textwidth}(a)\hspace{0.35\textwidth}
    \includegraphics[width=0.51\linewidth]{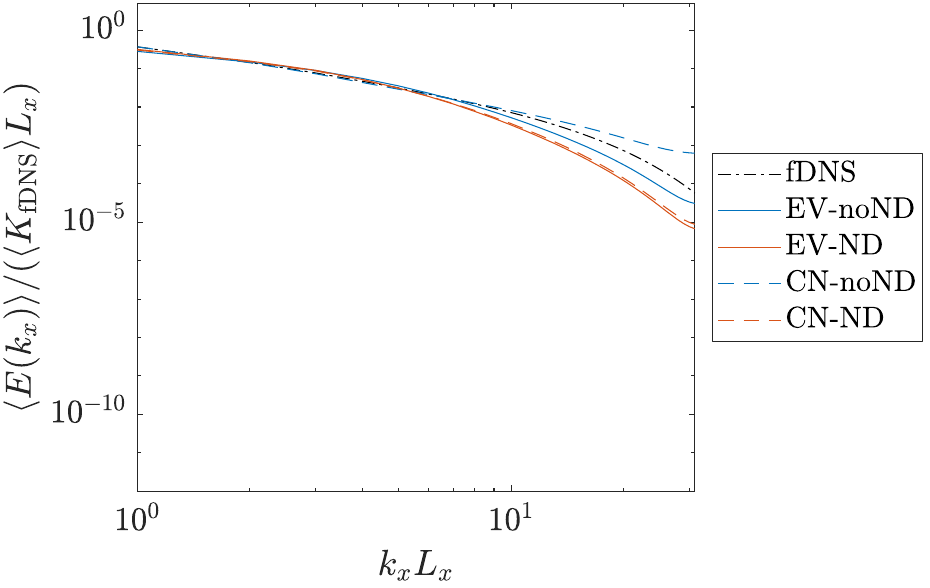}
    \hspace{-0.50\textwidth}(b)\hspace{0.48\textwidth}
    \caption{
    Energy spectra of \textit{a posteriori} analysis of data-driven constant coefficient eddy-viscosity (solid) and complex nonlinear (dashed) models, excluding (blue) and including (red) numerical deviation for
    filter width of (a) $\sigma/\Delta_{\textrm{LES}}=1.0$ and (b) $\sigma/\Delta_{\textrm{LES}}=0.5$, compared to filtered DNS (black dot-dashed). }
    \label{fig:apost-1Del}
\end{figure}

The impact of numerical deviation is further examined by generalizing the results to other numerical setups, beginning with the effect of filter size. Increasing the filter size relative to the grid spacing is known to reduce the relative contribution of numerical errors \cite{chow2003further}, and it has also been shown to improve the accuracy of data-driven models \cite{huang2024consistent}. We vary the filter width $\sigma/\Delta_{\textrm{LES}}$ across four values: 0.5, 1.0, 2.0, and 4.0.

For a smaller filter width, $\sigma/\Delta_{\textrm{LES}} = 1.0$, the \textit{a priori} evaluation (Table \ref{tab:apriori}) shows slight degradation compared to the baseline case of $\sigma/\Delta_{\textrm{LES}} = 2.0$. 
The corresponding \textit{a posteriori} results, shown in Fig. \ref{fig:apost-1Del}(a), indicate a familiar overprediction of small-scale turbulence, particularly in the complex nonlinear model. 
As seen previously, models trained without the numerical deviation again exhibit a mismatch between the \textit{a priori} and \textit{a posteriori} behaviors. 
Incorporating the numerical deviation during training suppresses this overprediction and {
yields an energy spectrum that is comparably closer to that of the filtered DNS.} 
Surprisingly, the smaller filter width leads to less small-scale overprediction compared to the baseline. 
Although numerical error may be more pronounced in the \textit{a priori} setting for smaller $\sigma$, the interaction between numerical and subfilter-scale terms reduces its impact in the \textit{a posteriori} tests.

Similarly, for an even smaller filter width $\sigma/\Delta_{\textrm{LES}} = 0.5$, Fig. \ref{fig:apost-1Del}(b) shows that including the numerical deviation continues to suppress small-scale overprediction. 
However, the overprediction is so minimal that both models—with and without the numerical deviation—provide a similarly accurate fit to the filtered DNS.

At the other extreme, for a larger filter width $\sigma/\Delta_{\textrm{LES}} = 4.0$, the \textit{a priori} accuracy remains comparable to the baseline (Table \ref{tab:apriori}). 
However, the \textit{a posteriori} simulations fail to sustain turbulence, suggesting that excessive filtering can overdamp the resolved motions, compromising model viability. 
This highlights the trade-off in choosing filter width: it must be large enough to reduce numerical contamination, yet small enough to preserve essential flow structures. 
In this context, including the numerical deviation during training aligns the \textit{a priori} and \textit{a posteriori} evaluations more consistently, with the greatest benefit seen for intermediate filter widths.

Next, we demonstrate the generalizability of the trained models with respect to different time step sizes. Since the model development is entirely local and time-independent, the implementation is not restricted to the time step used in the DNS-aided LES. 
In Figs. \ref{fig:apost-1term} and \ref{fig:apost-4terms}, the dashed and dotted lines represent simulations using time steps of $2\Delta t$ and $5\Delta t$, respectively. 
These results are nearly identical to those obtained with the original time step $\Delta t$, confirming that the trained models are robust to changes in temporal resolution.

\begin{figure}
    \centering
    \includegraphics[width=0.38\linewidth, trim={0 0 3.8cm 0}, clip]{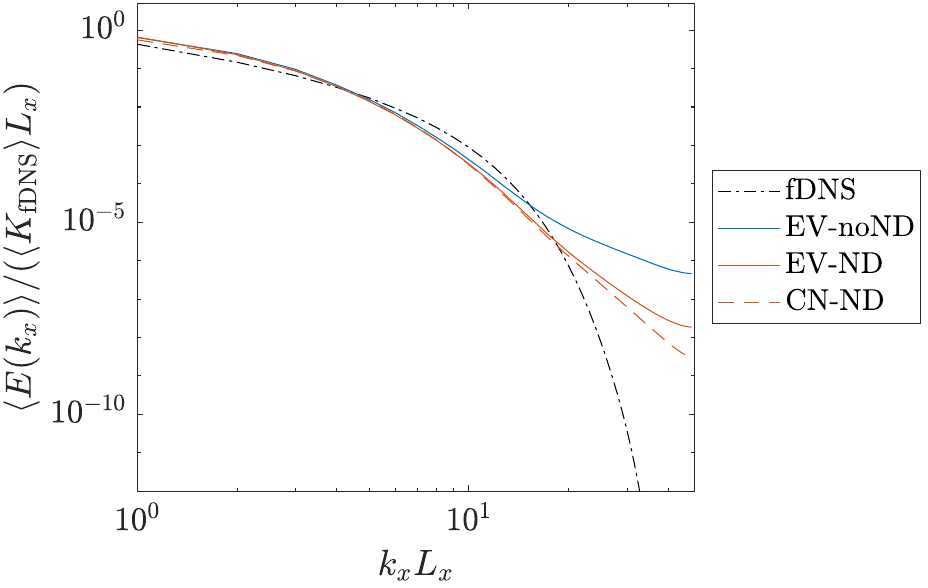} ~ 
    \hspace{-0.37\textwidth}(a)\hspace{0.35\textwidth}
    \includegraphics[width=0.51\linewidth]{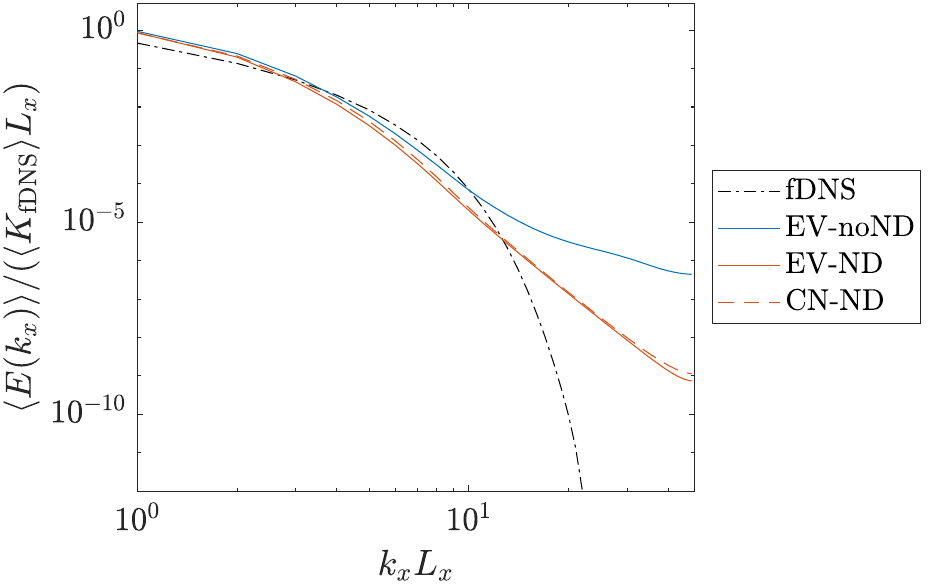}
    \hspace{-0.50\textwidth}(b)\hspace{0.48\textwidth}
    \caption{
    Energy spectra of \textit{a posteriori} analysis of data-driven constant coefficient eddy-viscosity (solid) and complex nonlinear (dashed) models, excluding (blue) and including (red) numerical deviation for
    filter width of (a) $\sigma/\Delta_{\textrm{LES}}=2.0$ and (b) $\sigma/\Delta_{\textrm{DNS}}=8.0$, compared to filtered DNS (black dot-dashed). }
    \label{fig:apost-96}
\end{figure}

Finally, we examine the impact of grid resolution on model performance. The same comparative analysis is conducted for LES at a finer resolution of $96^3$, and the results are compared with the baseline cases discussed in Sec. \ref{sec:results}. 
Two sets of tests are performed: one maintains a constant filter-to-grid ratio, $\sigma/\Delta_{\textrm{LES}} = 2.0$, while the other maintains a constant absolute filter width, $\sigma/\Delta_{\textrm{DNS}} = 8.0$. 
As shown in Fig. \ref{fig:apost-96}, the \textit{a posteriori} results at higher resolution reaffirm that including the numerical deviation leads to more consistent and stable LES behavior. 
In contrast, the complex nonlinear model trained without accounting for the numerical deviation becomes numerically unstable. 
This highlights that although complex nonlinear models offer greater flexibility in capturing subgrid-scale dynamics, they are also more prone to numerical instability when the numerical deviation is not included during training. 
Furthermore, grid refinement can destabilize an SGS model that was stable on a coarser grid. 
For both the eddy-viscosity and the complex nonlinear models, the influence of the numerical deviation becomes more pronounced at finer resolutions, as a greater range of small-scale turbulence is resolved in the flow field.

\section{Conclusion}
\label{sec:conclusion}


In this work, we investigate the inconsistency between the \textit{a priori} and \textit{a posteriori} tests in data-driven SGS modeling. While data-driven approaches often demonstrate improved accuracy during training (\textit{a priori} tests), their performance in actual LES simulations (\textit{a posteriori} tests) does not always surpass that of traditional models. Furthermore, data-driven models frequently suffer from numerical instability. This inconsistency arises because the governing equations underlying the training data—typically filtered DNS—and those governing the LES simulations differ due to commutation errors and resolution disparities. This discrepancy introduces a numerical deviation that can be locally significant compared to the SGS stress term \cite{huang2024consistent}.

To assess the impact of this numerical deviation on model development, we trained neural network SGS models both including and excluding the numerical deviation, considering two model forms: an eddy-viscosity and complex nonlinear models. 
{Although the \textit{a priori} performance of models including the numerical deviation may appear inferior to those excluding it, their \textit{a posteriori} kinetic energy spectra are generally comparable or even superior.} 
The nonlinear model captures richer small-scale turbulence behavior but may obscure large-scale structures. Models excluding the numerical deviation tend to significantly overpredict small-scale turbulence, resulting in severely distorted energy spectra. 
Incorporating the numerical deviation suppresses this overprediction, stabilizing the simulation and yielding energy spectra that align well with filtered DNS data. 
This effect is particularly critical for complex nonlinear models, which are essential for capturing anisotropic SGS behaviors.

The importance of accounting for numerical deviation is heightened in data-driven approaches due to their inherent nonlinearity. Our comparison of the mixed model and the data-driven nonlinear model with constant coefficients confirms that excluding the numerical deviation leads to exaggerated small-scale turbulence predictions when the model form allows it. Including the numerical deviation results in more physically consistent predictions. Although strategies such as increasing the filter-grid ratio and applying explicit filtering techniques can mitigate numerical errors, they do not fully resolve them. The model inconsistency problem is more challenging in data-driven models because nonlinear interactions cause errors to accumulate, causing the LES solution to drift away from the data-driven model’s valid state space. Thus, incorporating the numerical deviation is crucial for achieving consistency between \textit{a priori} and \textit{a posteriori} tests in data-driven SGS modeling.

We also extend the benefits of including the numerical deviation to different filter widths and LES resolutions. Unlike in non-data-driven methods, increasing the filter width does not necessarily reduce the numerical error term here. While including the numerical deviation generally improves model consistency, the model form must remain physically sound, and the filter width of the filtered DNS should be chosen at an intermediate scale to ensure satisfactory \textit{a posteriori} performance. Importantly, including the numerical deviation does not impose additional constraints on spatial or temporal discretization schemes. The trained model requires only local information and can be applied flexibly across different time steps and grid configurations.

\begin{acknowledgements}
This work is supported by the Center for Turbulence Research at Stanford University and the Office of Naval Research under Grant Number N00014-23-1-2729. 
The authors would like to thank Dr. Anthony Leonard for the discussion on the tensor diffusivity model. 
\end{acknowledgements}

\appendix*

\section{Training history}
\label{sec:history}

\begin{figure}
    \centering
    \includegraphics[width=0.5\linewidth]{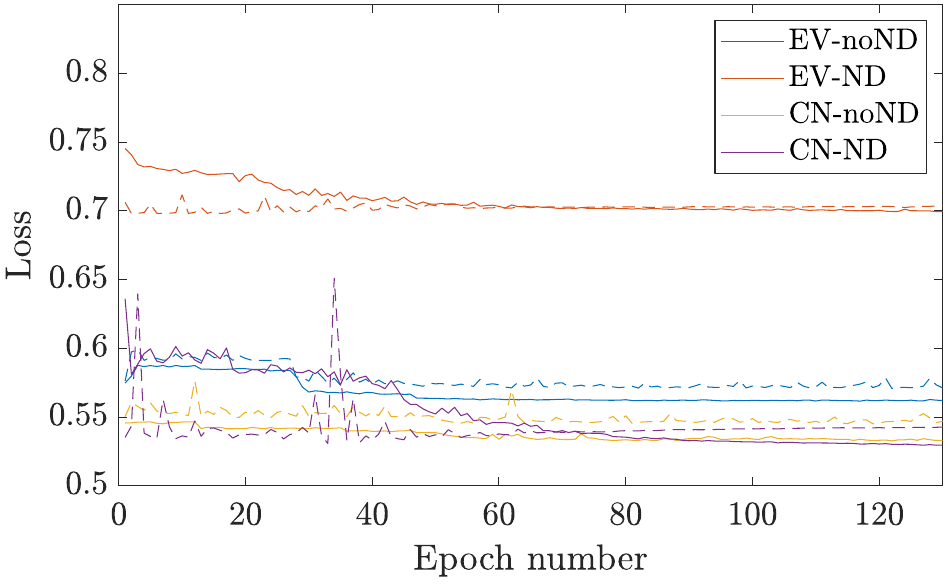} 
    \caption{
    The training history of the loss function values for the four models. Solid and dashed lines denote training and validation losses, respectively. EV and CN represent eddy-viscosity and complex nonlinear models; noND and ND indicate exclusion and inclusion of numerical deviation.}
    \label{fig:training-history}
\end{figure}

{
The training history of the examined models is shown in Fig. \ref{fig:training-history}, where the convergence of the loss function of the training dataset and that of the validation dataset indicates that the training is finished without overfitting. 
All four models have shown a converged performance after approximately 100 epochs.} 

\bibliographystyle{ieeetr}
\bibliography{eLES-ref}

\end{document}